\documentclass[preprint,amsmath,amssymb,nofootinbib]{revtex4}
\input epsf
\usepackage{epsfig}
\usepackage{amsmath}
  
 \def\d{\partial} \def\l{\left(} \def\r{\right)}
  \newcommand{\be}{\begin{equation}}
\newcommand{\ee}{\end{equation}} \newcommand{\bea}{\begin{eqnarray}}
\newcommand{\eea}{\end{eqnarray}} \newcommand{\bg}{\begin{gather}}
\newcommand{\eg}{\end{gather}}
\newcommand{\bseq}{\begin{subequations}}
\newcommand{\eseq}{\end{subequations}}

 \newcommand{\Tr}{{\rm
Tr}} \def\half{\frac{1}{2}}

\begin{document}
\title{Cosmological attractors in massive gravity}
\author{S.L.~Dubovsky$^{a,c}$} 
\author{P.G.~Tinyakov$^{b,c}$}
\author{I.I.~Tkachev$^{a,c}$} 
\affiliation{ $^a$Department of Physics, CERN
Theory Division, CH-1211 Geneva 23, Switzerland\\ $^b$Service de Physique
Th\'eorique, Universit\'e Libre de Bruxelles, CP225, blv.~du Triomphe, B-1050
Bruxelles, Belgium\\ $^c$Institute for Nuclear Research of the Russian Academy
of Sciences, 60th October Anniversary Prospect, 7a, 117312 Moscow, Russia }%

\begin{abstract}
We study Lorentz-violating models of massive gravity which preserve
rotations and are invariant under time-dependent shifts of the spatial
coordinates. In the linear approximation the Newtonian potential in
these models has an extra ``confining'' term proportional to the
distance from the source. We argue that during cosmological expansion
the Universe may be driven to an attractor point with larger symmetry
which includes particular simultaneous dilatations of time and space
coordinates.  The confining term in the potential vanishes as one
approaches the attractor. In the vicinity of the attractor the extra
contribution is present in the Friedmann equation which, in a certain
range of parameters, gives rise to the cosmic acceleration.
\end{abstract}
\maketitle

\section{Introduction}
Recently it has been
realized~\cite{Arkani-Hamed:2003uy,Rubakov:2004eb,Dubovsky:2004sg}
that models of modified gravity which contain Lorentz-violating
graviton mass terms may avoid all known problems of massive gravity
such as the van Dam-Veltman-Zakharov (vDVZ)
discontinuity~\cite{vanDam:1970vg,Zakharov}, ghost
instabilities~\cite{Boulware:1973my} and strong quantum effects at the
unacceptably low energy scale~\cite{Arkani-Hamed:2002sp}\footnote{It
is worth mentioning that another intriguing route to solve these
problems may be to take into account the effects of local
curvature~\cite{Vainshtein:1972sx,Deffayet:2001uk,Nicolis:2004qq}.}.
In the most general sense by massive gravity we understand any theory
described by the following action
\begin{equation}
\label{fullaction}
S=-M_{Pl}^2\int d^4x\sqrt{-g}R+\int d^4x\sqrt{-g}F\;, 
\end{equation} 
where the first term is a usual Einstein--Hilbert term and $F$ is,
generally speaking, an arbitrary (non-covariant) function of metric
components and their derivatives. Matter fields are assumed to be
covariantly coupled to the metric. A systematic study of the
rotationally invariant theories of massive gravity was performed in
Ref.~\cite{Dubovsky:2004sg}.  A particularly interesting class of
models found there is characterized by a residual reparametrization
symmetry
\begin{equation}
\label{symmetry}
x^i\to x^i+\xi^i(t)\;, 
\end{equation} 
where $x^i$ are the spatial coordinates. This class of models is
singled out by the following two properties.  First, in the vicinity
of the Minkowski background these models represent consistent
low-energy effective theories valid up to the energy scale
$\Lambda\sim\sqrt{mM_{pl}}$, where $m$ is a graviton mass.  The
absence of ghosts and classical instabilities is ensured by the
symmetry requirements alone without need for any extra fine-tuning.
The second important property is that already the lowest-dimension
operators (graviton mass terms) lead to the modification of gravity, in
particular gravitational waves are massive. An example of a theory
which shares the first but not the second property is the ghost
condensate model~\cite{Arkani-Hamed:2003uy} which, at the lowest
derivative level, is equivalent to the gauge-fixed Einstein
theory. Another class of theories with this property is discussed in
Refs.~\cite{Jacobson:2001yj,Gripaios:2004ms}. In what follows by
massive gravity we understand gravity theories (\ref{fullaction})
obeying symmetry (\ref{symmetry}).

The phenomenological consequences of the massive gravity models
obeying symmetry (\ref{symmetry}) were first studied in
Ref.~\cite{Dubovsky:2004ud}. One of the unexpected properties found
there is that massive gravitational waves may coexist with the
long-range potential between the sources.  In general, the
gravitational potential in these models contains an extra
``confining'' piece which grows linearly with the distance from the
source,
\begin{equation}		
\label{confpotential}
\Phi=G_N M \left( -{1\over r}+ \mu^2 r\right),
\end{equation} 
where $\mu^2$ is a combination of the graviton masses (see
Sect.~\ref{sect:linear}) proportional to their overall scale.  The
analysis of Ref.~\cite{Dubovsky:2004ud} was focused mainly on the case
$\mu^2=0$ when the additional dilatation symmetry ensures that the
long-range potential is identical to that of the Einstein theory. This
relaxes the constraints coming from the Solar system and
Cavendish-type experiments, and opens up a possibility for the
graviton mass to be as large as $\sim (10^{-15}\mbox{cm})^{-1}$
without contradiction to the existing experimental data. The relic
gravitational waves produced at inflation may constitute today the
cold dark matter in the Universe and would give a unique
monochromatic signal in the gravitational wave detectors
\cite{Dubovsky:2004ud}.

The purpose of the present paper is to study cosmological solutions in
the massive gravity. We address the question of whether such solutions
are phenomenologically acceptable, and what are their generic
properties. We do not assume that parameters are tuned so that
$\mu^2=0$ from the very beginning. It turns out, however, that for a
large class of functions $F$ in Eq.~(\ref{fullaction}) the cosmological
evolution naturally drives the system to the point $\mu^2=0$ where
there are no corrections to the Newtonian potential. In other words,
this point is an attractor in the solution space (it corresponds to
the restoration of an additional dilatation symmetry). At this point
the graviton mass has a finite non-zero value, while the modification
of the Friedmann equation has a form of an extra term which behaves
like a dark energy with the equation of state depending on the
parameters of the model (properties of the function $F$).

The paper is organized as follows. We begin in Sect.~\ref{sect:linear}
by analyzing linear perturbations about the flat background and, in
particular, derive Eq.~(\ref{confpotential}). In Sect.~\ref{cosmology}
we study general properties of the cosmological solutions in massive
gravity.  We then consider in Sect.~\ref{stability} the stability of
the curved solutions against perturbations of high momenta (the
Boulware-Deser instability) and argue that the stability is achieved
without fine-tuning of parameters at least for backgrounds close to
the Minkowski space. In particular, cosmological solutions found in
Sect.~\ref{cosmology} are stable in the vicinity of the attractor
point. In the concluding Sect.~\ref{discussion} we discuss possible
phenomenological applications and some future directions in the
studies of the massive gravity.

\section{Linearized theory near Minkowski background}
\label{sect:linear}

As has been argued in Ref.~\cite{Dubovsky:2004sg}, a convenient way to
describe the Lorentz-violating models of massive gravity is to
introduce the set of four scalar ``Goldstone'' fields $\phi^0$,
$\phi^i$, which have a particular derivative couplings to gravity.  In
terms of the metric and the Goldstone fields, the action has a
generally-covariant form. The spontaneous breaking of the covariance
is achieved by assuming non-zero vacuum expectation values of the
derivatives of the Goldstone fields.  The Goldstone fields can be
eliminated from the action by a suitable coordinate transformation; in
such a ``unitary gauge'' the action only depends on the metric
components.

A class of Lorentz-violating gravity models which possess the symmetry
(\ref{symmetry}) and rotational invariance is represented by the action
\begin{equation}
\label{action}
S=\int d^4x\sqrt{-g}\left[ - M_{Pl}^2
R+\Lambda^4F(X,W^{ij},\dots)\right] ,
\end{equation}
where
\[
X=g^{\mu\nu}\d_\mu\phi^0\d_\nu\phi^0,
\]
\begin{equation}
W^{ij}={g^{\mu\nu}\d_\mu\phi^i\d_\nu\phi^j}-
{g^{\mu\nu}\d_\mu\phi^0\d_\nu\phi^i\cdot
g^{\lambda\rho}\d_\lambda\phi^0\d_\rho\phi^j\over X}  \;,
\label{XVYW}
\end{equation}
dots stand for higher derivative terms, and $\Lambda$ is a parameter
which determines the cutoff scale of the theory. The indices $i$, $j$
are converted using $\delta_{ij}$. Low energy modification of gravity
takes place at the scale $m\sim \Lambda^2/M_{Pl}$. The function $F$ is
arbitrary apart from the constraints following from the requirement
that the model is free of ghosts and strong-coupling problems
\cite{Dubovsky:2004sg}; we assume that it depends on a
single scale $\Lambda$. The coefficient in front of the
Einstein-Hilbert action is chosen for convenience.

We assume that the model (\ref{action}) possesses the solution which
corresponds to the Minkowski space,
\begin{eqnarray}
\nonumber
g_{\mu\nu} &=& \eta_{\mu\nu},\\
\nonumber
\phi^0 &=& \alpha \Lambda^2t,\\
\label{vacuum}
\phi^i &=& \beta \Lambda^2x^i. 
\end{eqnarray}
Here $\alpha$ and $\beta$ are some constants which have to be chosen
in such a way that the energy-momentum tensor of the Goldstone fields
is zero. This requirement reduces to two equations (\ref{constraint1})
and (\ref{constraint2}), shown in the Appendix A.  Consequently, this
choice is possible for a generic function $F$.

Our current goal is to study linear perturbations about the vacuum
(\ref{vacuum}). It is convenient to work in the ``unitary gauge'' where
the Goldstone fields are set to their vacuum values (\ref{vacuum}). In
this gauge the remaining perturbations are those of the metric,
$\delta g_{\mu\nu}$,
\begin{equation}
g_{\mu\nu} = \eta_{\mu\nu}+ \delta g_{\mu\nu}.
\label{delta_g}
\end{equation}
Following the notations of Ref.~\cite{Mukhanov:1990me}, we parameterize
$\delta g_{\mu\nu}$ as follows, 
\begin{eqnarray}
\nonumber
\delta g_{00} &=& 2\varphi;\\
\nonumber
\delta g_{0i} &=&  S_i - \d_i B;\\
\nonumber
\delta g_{ij} &=& - h_{ij} - \d_i F_j - \d_j F_i 
+ 2 (\psi \delta_{ij} - \d_i\d_j E), 
\end{eqnarray}
where $h_{ij}$ are the transverse and traceless tensor perturbations,
$S_i$ and $F_i$ are the transverse vector perturbations, while $\varphi$,
$\psi$, $B$ and $E$ are the scalar perturbations. The potential $\varphi$
is not to be confused with the Goldstone fields $\phi_0$ and
$\phi_i$.

The quadratic Lagrangian for perturbations has the form 
\begin{equation}
L = L_{EH} + L_m + L_s,
\label{quadratic_action}
\end{equation}
where the three contributions are the Einstein-Hilbert term, the mass
term and the source term, respectively. The quadratic part of the
Einstein-Hilbert Lagrangian is
\[
L_{EH}=  M_{Pl}^2\Bigl\{
- {1\over 4} h_{ij} (\d_0^2-\d_i^2)h_{ij} 
- {1\over 2} (S_i + \d_0F_i)\d_j^2(S_i + \d_0F_i)
\]
\begin{equation}
+ 4(\varphi+\d_0B-\d_0^2E) \d_i^2\psi + 6\psi\d_0^2\psi
-2 \psi\d_i^2\psi \Bigr\}. 
\label{EH}
\end{equation}
The mass term originates from the second term in Eq.~(\ref{action}).
We parameterize the mass parameters according to the notations of
Ref.~\cite{Rubakov:2004eb}, 
\begin{gather}
\label{masterm1}
L_m = {M_{Pl}^2\over 4} 
\Bigl\{ m_0^2 \delta g_{00}^2 
+ 2 m_1^2 \delta g_{0i}^2 - m_2^2 \delta g_{ij}^2 
+ m_3^2 \delta g_{ii}^2 - 2 m_4^2 \delta g_{00} \delta g_{ii} \Bigr\}.
\end{gather}
The contribution proportional to $m_1^2$ is absent in our model. This
is guaranteed by the symmetry (\ref{symmetry}). In terms of tensor,
vector and scalar perturbations these mass terms read 
\begin{gather}
M_{Pl}^2\Bigl\{ -{1\over 4} m_2^2 h_{ij}^2 
-  {1\over 2} m_2^2 (\d_i F_j)^2 
+ m_0^2\varphi^2 + \l m_3^2-m_2^2\r\l\d_i^2E\r^2-\nonumber\\
- 2(3m_3^2-m_2^2)\psi\d_i^2E + 3\l 3m_3^2-m_2^2\r\psi^2+
2m_4^2\varphi\d_i^2 E - 6m_4^2\varphi\psi \Bigr\} .
\label{massterm2}
\end{gather}
The masses $m_i^2$ are expressed in terms of the first and second
derivatives of the function $F$, the parameter $\Lambda$ and the
Planck mass. The overall scale of masses is set by the ratio
$\Lambda^2/M_{Pl}$.  The explicit expressions are given in the
Appendix A.

To probe the linear response of the system we add the source
$T_{\mu\nu}$ which is assumed to be conserved,
$\d^{\mu}T_{\mu\nu}=0$. The corresponding contribution to the
Lagrangian can be written as
\[
L_s=-T_{00}\l\varphi+\d_0B-\d_0^2E\r-T_{ii}\psi + (S_i + \d_0F_i)T_{0i} 
+ {1\over 2} h_{ij} T_{ij}.
\]
All combinations coupled to the components of $T_{\mu\nu}$ are
gauge-invariant. The one multiplying $T_{00}$,
\[
\Phi\equiv\varphi+\d_0B-\d_0^2E,
\] 
plays the role of the Newtonian potential in the non-relativistic limit.

\paragraph{Tensor sector.} In the tensor sector, only the transverse
traceless perturbations $h_{ij}$ are present. Their field equation
is that of a massive field with the mass $m_2$, in agreement with
Ref.~\cite{Rubakov:2004eb}. Thus, there are two massive spin 2
propagating degrees of freedom.

\paragraph{Vector sector.} In the vector sector, the field
equations read 
\begin{gather}
\label{dSi1}
- \d_j^2 (S_i + \d_0 F_i)=- T_{0i},\\
\label{dSi2}
\d_0\d_j^2 ( S_i + \d_0F_i) + m_2^2 \d_j^2 F_i = \d_0 T_{0i}.
\end{gather}
Taking the time derivative of Eq.~(\ref{dSi1}) and adding it to 
Eq.~(\ref{dSi2}) gives
\[
F_i =0,
\]
provided that $m_2^2\neq 0$.  Thus, the vector sector of our model
behaves in the same way as in the Einstein theory in the gauge
$F_i=0$.  There are no propagating vector perturbations and gravity is
not modified in the vector sector unless one takes into account
non-linear effects or higher derivative terms.

\paragraph{Scalar sector.} The field equations for scalar
perturbations  are
\begin{eqnarray}
\label{psieq}
&&2\d_i^2\psi+m_0^2\varphi+m_4^2\d_i^2E-3m_4^2\psi={T_{00}\over 2M_{Pl}^2},\\
\label{phieq}
&&2\d_i^2\Phi-2\d_i^2\psi+6\d_0^2\psi-
\l 3m_3^2-m_2^2\r\d_i^2E+3\l 3m_3^2-m_2^2\r\psi-3m_4^2\varphi=
{T_{ii}\over 2M_{Pl}^2},\\
\label{Eeq}
&&-2\d_i^2\d_0^2\psi+\l m_3^2-m_2^2\r\d_i^4 E-\l 3m_3^2-m_2^2\r\d_i^2\psi+
m_4^2\d_i^2\varphi=-{\d_0^2T_{00}\over 2M_{Pl}^2},\\
\label{Beq}
&&2\d_i^2\d_0\psi={\d_0T_{00}\over 2M_{Pl}^2}.
\end{eqnarray}
Eq.~(\ref{Beq}) implies 
\begin{equation}
\psi={1\over \d_i^2}{T_{00}\over 4M_{Pl}^2}+\psi_0(x^i),
\label{psi}
\end{equation}
where $\psi_0(x^i)$ is some time-independent function.

From Eqs.~(\ref{psieq}) and (\ref{Eeq}) one finds
\begin{eqnarray}
\label{phi}
&&\varphi={2m_2^2m_4^2\over \Delta}\psi + {2(m_3^2-m_2^2)\over \Delta} 
\d_i^2\psi_0, \\
\label{E}
&&\d_i^2E=  \l 3 - {2m_0^2m_2^2\over \Delta}\r \psi 
- {2m_4^2\over \Delta} \d_i^2 \psi_0,
\end{eqnarray}
where 
\[
\Delta = m_4^4-m_0^2(m_3^2-m_2^2).
\]
Finally, substituting Eqs.~(\ref{psi}), (\ref{phi}) and (\ref{E}) into
Eq.~(\ref{phieq}) one finds the gauge-invariant potential $\Phi$,
\begin{gather}
\label{Phi}
\Phi={1\over \d_i^2} {T_{00}+ T_{ii}\over 4M_{Pl}^2} 
- 3{\d_0^2\over \d_i^4}{T_{00}\over 4 M_{Pl}^2} 
+  \l 3 - { 2m_0^2 m_2^2\over
\Delta} \r {m_2^2\over \d_i^2} 
\l{1\over \d_i^2}{T_{00}\over 4 M_{Pl}^2} + \psi_0 \r
+ \l 1 -{2m_2^2m_4^2\over \Delta }\r \psi_0,
\end{gather}
where we presented explicitly the dependence on $\psi_0$ and
$T_{\mu\nu}$.  The first two terms on the r.h.s. of Eq.~(\ref{Phi})
are the standard contributions in the Einstein theory, the first
becoming the Newtonian potential in the nonrelativistic limit. Thus,
barring the $\psi_0$-dependent terms, the gauge-invariant potentials
$\Phi$ and $\psi$ in our model differ from their analogs in the
Einstein theory $\Phi_E$ and $\psi_E$ by the mass-dependent third term
on the r.h.s of Eq.~(\ref{Phi}),
\begin{eqnarray}
\nonumber
&&\psi=\psi_E,\\
\label{PPhi}
&&\Phi=\Phi_E + \l 3 - {2m_0^2m_2^2 \over\Delta} \r
{m_2^2\over \d_i^4}
{T_{00}\over 4M_{Pl}^2}.
\end{eqnarray}
This term vanishes if all masses uniformly go to zero, which implies
the absence of the vDVZ discontinuity.  Eq.~(\ref{PPhi}) is the result
presented in Ref.~\cite{Dubovsky:2004ud}.  For a static source,
Eq.~(\ref{PPhi}) leads to the modification of the Newtonian potential
of a point mass $M$ as shown in Eq.~(\ref{confpotential}) with
\begin{equation}
\mu^2 = -{1\over 2} m_2^2\l 3 - {2 m_0^2 m_2^2 \over\Delta}\r.
\label{mu}
\end{equation}
This indicates the breakdown of perturbation theory at distances
$r\gtrsim 1/(G_NM\mu^2)$.  Note, that the modification of the
Newtonian potential is absent if $3\Delta=2m_0^2m_2^2$ (and
$\Delta\neq 0$). We will see in Sect.~\ref{cosmology} that this
happens in the vicinity of the cosmological attractor, i.e. at late
times of the cosmological evolution.

The freedom of choosing the time-independent function $\psi_0(x)$,
which enters the above gravitational potentials, indicates the
presence of the scalar mode with the dispersion relation\footnote{In
what follow $\omega$ denotes frequency and $p$ denotes the absolute
value of the spatial momentum.} $\omega^2=0$ \cite{Dubovsky:2004sg}
(cf. also Sect.~\ref{stability}, Eq.~(\ref{velocity}) in the limit
$m_1^2\to 0$). This mode is an analogue of the ghost condensate
mode~\cite{Arkani-Hamed:2003uy} and becomes dynamical with the account
of higher-derivative terms in the action (\ref{action}), acquiring the
dispersion relation $\omega^2\propto p^4$ (so that $\psi_0$ becomes a
slowly-varying function of time).  The value of $\psi_0$ is fixed by
the initial conditions.  In the linear regime, the non-zero value
$\psi_0$ would mean the presence of the incoming ``ghost condensate
wave''.  So, a physically reasonable choice of $\psi_0$ is
$\psi_0(x^i)=0$.  We will discuss in more detail a possible role of
this mode in the concluding Sect.~\ref{discussion}.

Note that the kinetic term of the ghost condensate mode is
proportional to the combination $\Delta$~\cite{Dubovsky:2004sg} (see
also Sect.~\ref{stability}). Therefore, in general this combination
should be non-zero (positive), in agreement with
Eqs.~(\ref{phi})---(\ref{PPhi}).  It may happen that in some special
cases one can obtain a healthy theory even if $\Delta=0$. An
interesting possibility suggested recently in
Ref.~\cite{Gabadadze:2004iv} is to impose an additional condition
$m_4^2=m_0^2=0$. Then the ``ghost condensate'' mode does not appear in
the linearized theory at the lowest derivative level in flat
background. It acquires both kinetic and gradient term at a
higher-derivative level, so that an additional symmetry $t\to
t+\xi^0(t)$ is needed to prevent this mode from being a ghost. The
gravitational potentials are non-singular in this case despite
$\Delta=0$ and have the same structure as our Eq.~(\ref{PPhi}).  As we
discuss in Sect.~\ref{stability}, the stability of this model requires
further study and is more subtle than in the case $\Delta\neq 0$
because of new propagating modes that appear in curved backgrounds.

\section{Cosmological solutions}
\label{cosmology}

Let us discuss flat cosmological solutions in the theory defined by
the action (\ref{action}). The flat cosmological ansatz is
\begin{gather}
\label{simple_ansatz}
ds^2= dt^2-a^2(t)dx_i^2,\\
\phi^0=\phi(t)\;,\;\;
\phi^i=\Lambda^2x^i\;.
\end{gather}
For this ansatz $W^{ij}=-a^{-2}\delta^{ij}$, so the function $F$ in
(\ref{action}) depends only on $X$ and $a$, $F = F(X,a)$. The Einstein
equations reduce to the Friedmann equation (see Appendix A),
\begin{gather}
\label{00eq}
\l{\dot{a}\over a}\r^2=
{1\over 6M_{Pl}^2}\Bigl\{ \rho_m +  2\Lambda^4X F_X 
- \Lambda^4F\Bigr\}
\equiv {1\over 6M_{Pl}^2}\Bigl\{ \rho_m +  \rho_1 + \rho_2\Bigr\}\;,
\end{gather}
where $\rho_m$ is the energy density of ordinary matter not including
Goldstone fields, and the field equation for $\phi^0$,
\begin{equation}
\label{phi0}
\d_t\l {a^3 \sqrt{X} F_X}\r=0\;.
\end{equation}
It is straightforward to solve this system of equations for any given
function $F(X,a)$. After the integration, Eq.~(\ref{phi0}) gives an
algebraic equation which determines $X$ as a function of the scale
factor $a$. The dependence $X(a)$ as found from Eq.~(\ref{phi0})
determines the behavior of the Goldstone energy density
$\rho_1+\rho_2$ as a function of $a$.  This makes Eq.~(\ref{00eq}) a
closed equation for the scale factor $a(t)$.

From the point of view of cosmological applications, of particular
interest are solutions where the scale factor $a(t)$ goes to infinity
at late times. Since the graviton masses are linear combinations of
the function $F(X,a)$ and its derivatives, one may wonder whether they
remain finite or go to zero in this limit, and whether the
effective-theory description remains valid. Indeed, Eq.~(\ref{phi0})
implies that at late times either $X$ or $F_X$ go to zero.  If $X\to
0$, then the expressions given in the Appendix A suggest that the
graviton masses go to zero as well. This may lower the cutoff scale of
the effective theory. Similarly, some of the masses apparently vanish
if $X$ goes to a finite value $X_0$ such that $F_X(X_0,a)\to0$. If $X$
goes to infinity, this questions the validity of the low-energy
effective theory by itself.

Let us show that, in spite of the naive expectations, for a wide class
of functions $F$ there exist solutions for which graviton masses are
finite in the limit $a\to\infty$ and the effective theory description
remains valid.  Assume that $X(a)$ asymptotes to some power of $a$ at
large $a$. This is not a very restrictive assumption --- for instance,
it is satisfied for any algebraic function $F(X,a)$.  Then there
exists such $\gamma$ that the combination $X^\gamma/a^2$ goes to a
non-zero constant as $a\to\infty$.  Eq.~(\ref{phi0}) implies that
$XF_X=\mbox{const}\cdot \sqrt{X}/a^3$; this determines the dependence
of the energy component $\rho_1$ on the scale factor,
\begin{equation}
\rho_1 = 
{\rm const} 
{1\over a^{3-{1/\gamma}}}.
\label{rho1}
\end{equation}
This relation generalizes the behavior found in the ghost condensate
models where the energy density of the ghost condensate scales like
$1/a^3$ (in our model the latter behavior is recovered at
$\gamma\to\infty$).

For $\gamma>1/3$ the energy density $\rho_1$ behaves like the dark
energy component with the the negative pressure.  Its equation of
state varies between that of the cold dark matter, $w=0$ (for
$\gamma=+\infty$), and that of the cosmological constant, $w=-1$ (for
$\gamma={1\over 3}$).  For $0<\gamma<1/3$ the term $\rho_1$ grows with
$a$. It corresponds to the energy density component with a highly
negative equation of state, $w<-1$. Without fine-tuning this
contribution cannot be canceled by the term $\rho_2$, so that the
Hubble rate diverges as $a\to\infty$ leading to the breakdown of the
low-energy effective theory and suggesting the presence of rapid
instabilities.  In what follows we assume that
$\gamma$ does not belong to this range. For $\gamma<0$ the energy
density $\rho_1$ corresponds to a fluid with a positive pressure.

In order to see that the graviton masses remain finite and the
effective field theory description is valid in the limit $a\to\infty$,
it is convenient to replace $X$ by a new variable $Z=X^\gamma/a^2$.
The function $F(X,a)$ becomes the function of $Z$ and $a$, $\tilde
F(Z,a)= F(Z^{1/\gamma}a^{2/\gamma},a)$. Note that it satisfies the
relation $\gamma Z \tilde F_Z = XF_X$, where $\tilde F_Z = \d \tilde
F/\d Z$.  In these notations Eq.~(\ref{phi0}) reads
\begin{equation}
\gamma a^{3-{1\over \gamma}} Z^{1-{1\over 2\gamma}} \tilde F_Z(Z,a) = A, 
\label{F_Z}
\end{equation}
where $A$ is an integration constant. This equation determines $Z$ as
a function of $a$. By construction, this dependence is such that
$Z(a\to\infty) = Z_0$, where $Z_0$ is some constant. 

If one assumes further that the function $\tilde F(Z,a)$ is regular at
$a\to\infty$, then at late times one has
\begin{equation}
F(X,a) = \tilde F(Z,a) \to F_0(Z). 
\label{FofZ}
\end{equation}
In terms of the original variables this means that in the limit
$a\to\infty$ the function $F(X,W^{ij})$ depends only on the
combination $X^\gamma W^{ij}$. This corresponds to the following
dilatation symmetry of the Goldstone action,
\begin{eqnarray}
\nonumber
\phi_0 &\to& \lambda \phi_0,\\
\phi_i &\to& \lambda^{-\gamma} \phi_i.
\label{dilatation}
\end{eqnarray}
In this case one has 
\[
\rho_2 = - \Lambda^4 F_0(Z_0),
\]
which behaves like a cosmological constant (assuming $F_0(Z_0)\neq
0$).  Likewise, at $a\to\infty$  the masses given by
Eqs.~(\ref{m0})--(\ref{m4}) become functions of $Z_0$ and in general
remain finite. 

In models with this kind of behavior of $X(a)$ the effective field
theory description remains valid even at $X\gg\Lambda^4$ provided
the value of $Z$ is small. This is guaranteed by the dilatation
symmetry (\ref{dilatation}) which relates configurations with
different values of $X$. Thus, there exists a wide class of functions
$F$ for which indefinitely expanding cosmological solutions are
compatible with constant graviton masses and allow for the effective
field theory description. 

Our assumptions about the function $F$ can be summarized in the
following expansion,
\begin{equation}
\label{epsilon}
F(Z,W)=F_0(Z)+\sum_{\nu>0}\epsilon^\nu W^{\nu}F_\nu(Z)\;,
\end{equation}
where $\nu$ takes positive (not necessarily integer) values,
$F_\nu(Z)$ are some regular functions of $Z$ (for shortness we have
suppressed the indices $i,\,j$) and $\epsilon$ is a formal expansion
parameter.  Eq.~(\ref{F_Z}) implies that an attractor point $Z_0$ is
determined by the condition $F_0'(Z_0)=0$, where prime denotes
$d/dZ$. Note that the expansion (\ref{epsilon}) does not need to hold
for arbitrary values of $Z$ and $W$; it is sufficient if it is
satisfied in some finite region around the attractor point.  One may
wonder whether the class of functions of the form (\ref{epsilon}) is
stable under quantum corrections. To see that this is generically the
case note that the action (\ref{epsilon}) is formally invariant under
the symmetry (\ref{dilatation}) provided one treats $\epsilon$ as a
spurion field transforming as
\[
\epsilon\to\lambda^{-2}\epsilon\;.
\]
Let us assume $\epsilon$ to be somewhat smaller than unity, so that
one can perform perturbation theory in this parameter. Then the
general form (\ref{epsilon}) of the action is invariant under quantum
corrections whenever expansions in $\epsilon$ works (i.e., no terms
proportional to negative powers of $\epsilon$ appear due to quantum
corrections).

The models with the function $F$ obeying Eq.~(\ref{FofZ}) have an
interesting feature which is a consequence of the symmetry
(\ref{dilatation}). It is straightforward to check that
Eq.~(\ref{dilatation}) implies the following relations among graviton
masses in the Minkowski space,
\begin{equation}
m_0^2= - 3\gamma m_4^2, \qquad \gamma(m_2^2 - 3 m_3^2) = m_4^2. 
\label{relation-for-masses}
\end{equation}
These relations ensure that the parameter $\mu^2$ defined by
Eq.~(\ref{mu}) is zero, i.e., the correction to the Newtonian
potential (the last term in Eq.~(\ref{PPhi})) vanishes.  Thus, barring
the effects of the higher derivative terms, at late times the only
modification of gravity at the linear level is the non-zero mass of
the two polarizations of the graviton. This suggests that the
confining term in the Newtonian potential is unlikely to have any
effect at present epoch. Indeed, the expressions for the graviton
masses given in the Appendix A imply that the correction to the
Newtonian potential goes to zero as $1/a^{2\nu_m}$, where $\nu_m$ is a
minimal value of $\nu$ in Eq.~(\ref{epsilon}).  This parameter has to
be fine-tuned to an extremely small value to allow for a substantial
value of $\mu^2$ at present.

A particularly simple case occurs when the function $F$ depends only
on the combination $Z= X^\gamma W^{ij}$.  If $\gamma > 1/3$ or
$\gamma<0$, the evolution drives the system to the point $\tilde
F_Z=0$, in full similarity with the ghost condensate model. In the
case $0<\gamma < 1/3$ and regular $\tilde F$, $Z$ has to diverge at
large $a$. This breaks the validity of the low energy effective
theory.

There are three boundary values of $\gamma$, which are somewhat
special, namely $\gamma=1/3,0,\infty$. If $\gamma=1/3$ then $Z$ is
constant during cosmological evolution and both $\rho_1$ and $\rho_2$
behave like a cosmological constant.  An interesting property of this
model is that a (constant) acceleration rate of the cosmological
expansion is determined by the initial conditions in the Goldstone
sector (the value of $Z$) rather than by the parameters of the action.

If $\gamma=0$ then $F(X,W)$ does not depend on $X$ at all.  In this
case Eq.~(\ref{phi0}) is satisfied automatically, and the only
unconventional component in the Friedmann equation (\ref{00eq}) is the
last term $\rho_2$. This term may describe arbitrary equation of state
depending on the choice of the function $F_0(W)$.  For functions $F$
regular when $W$ goes to zero, this term becomes a cosmological
constant as before.

In the case $\gamma=\infty$ the function $F$ depends on the scalar
quantities $X$, $\Tr W^2/(\Tr W)^2$ and $\Tr W^3/(\Tr W)^3$. Flat
cosmological solutions in such a theory have the same properties as in
the ghost condensate model where the $F$ is a function of $X$
only. These theories, however, differ from the ghost condensate model
in that they describe massive gravitons, and have different solutions
in a non-flat case.

It is worth commenting on the role of the regular Minkowski vacua
which are the points in the $(X,W)$ space at finite (non-zero) values
of $X$ and $W$ where the energy-momentum tensor of matter and
Goldstone fields is zero and thus the Minkowski metric solves the
Einstein equations. In the absence of matter, $\rho_m=0$, these points
are determined by Eqs.~(\ref{constraint1}) and
(\ref{constraint2}). There may exist solutions to Eqs.~(\ref{00eq})
and (\ref{phi0}) which asymptotically approach these points. These
solutions correspond to the scale factor going to a finite limit, so
they do not describe the current phase of the cosmological expansion.

To conclude this section we would like to stress that our analysis may
not exhaust all viable cosmological solutions in the model with the
action (\ref{action}).  For instance, the combination $X^\gamma/a^2$
may be proportional to some power of $\log a$ at late times, which is
the case not covered above. Another possibility is to consider more
general cosmological ansatz than that given by
Eq.~(\ref{simple_ansatz}). Namely, one may consider the time-dependent
configuration\footnote{We thank S.~Sibiryakov
for pointing out to us this possibility.} of the fields $\phi^i$,
\[
\phi^i=\Lambda^2 C(t)x^i,  
\]
where $C(t)$ is an arbitrary function of time. Due to the symmetry
(\ref{symmetry}) this ansatz is still homogeneous as the constant
shift of the spatial coordinates $x^i$ can be compensated by the
$\phi^0$-dependent shift of the fields $\phi^i$. The equations of
motion on this ansatz reduce to two equations: the Friedmann equation
(\ref{00eq}) which remains unchanged, and the equation for $\phi^0$, 
\begin{equation}
\label{phi0gen}
{X^{1/2}\over a^3}\d_t\l a^3 X^{1/2}F_X\r+3{\dot C\over C}WF_W=0.
\end{equation}
For any fixed function $C(t)$, Eqs.~(\ref{00eq}) and (\ref{phi0gen})
determine the dependence of $X$ and $a$ on time. Interestingly, in the
case when the function $F$ is invariant under the additional
dilatation symmetry (\ref{dilatation}), Eq.~(\ref{phi0gen}) takes the
form (\ref{F_Z}) irrespectively of the particular shape of
$C(t)$. Thus, while the time dependences of $X$ and $W$ separately
vary with the choice $C(t)$, the evolution of $Z$ and the scale factor
$a$ is universal.  Consequently, observable quantities such as the
expansion rate and the graviton masses do not depend on the function
$C(t)$ if the symmetry (\ref{dilatation}) holds.

The situation is different if the dilatation symmetry is absent: in
general, the expansion rate depends on the choice of the function
$C(t)$. This ambiguity is a consequence of the symmetry
(\ref{symmetry}) and is related to the presence of modes with the
dispersion relation $p_i^2=0$. In order to fix this ambiguity one
should specify boundary conditions for the fields $\phi^i$ at spatial
infinity. To see this, imagine that the space is compact. For
instance, if the space is a torus of the size $L$, the fields $\phi^i$
have to satisfy some kind of (quasi)periodicity condition, e.g.
\[
\phi^i(x^i)=\phi^i(x^i+L)-\Lambda^2L .
\]
This condition implies $C={\rm const}$. Other boundary conditions may
lead to time-dependent $C(t)$. In this sense, the ambiguity in choosing
different functions $C(t)$ is analogous to the ambiguity in choosing
the vacuum state in theories with flat directions.

\section{Stability}
\label{stability}

Let us discuss the stability of the cosmological solutions obtained in
the previous section. One should distinguish two different types of
instabilities which may occur in a theory which is
perturbatively stable about the flat background when the latter
becomes curved. The first type of instabilities has the characteristic
wavelength and time-scales much longer that the inverse cutoff scale
$\Lambda^{-1}$.  They are set either by the curvature of the
background, or suppressed by the powers of $\Lambda/M_{Pl}$, if these
instabilities appear due to mixing of higher derivative terms with
gravity (the latter type of instability is present, e.g., in the ghost
condensate) . We call these the infrared (IR) instabilities. Depending
on a particular situation, the IR instabilities, if present, may be
either dangerous or interesting phenomenologically (like, e.g., the
Jeans instability). Their analysis is clearly important for the
conclusion on the phenomenological viability of the model. However,
even if present, the IR instabilities do not question the
applicability of the analysis based on the low-energy effective field
theory.  We do not address IR stability of our models in the present
paper.

The instabilities of a different type, which we refer to as
ultraviolet (UV) instabilities, are those which occur at wavelengths
(and/or timescales) much shorter than that of the background
curvature, approaching the scales of order $\Lambda^{-1}$. Such
instabilities do affect the structure of the theory in the ultraviolet
and imply the breakdown of the effective field theory description for
scales much lower than $\Lambda$. An example of such an instability is
the Boulware--Deser instability~\cite{Boulware:1973my} which occurs in
the curved background in the Fierz--Pauli theory of massive gravity
due to the presence of the ghost mode\footnote{Note, however, that the
statement~\cite{Gabadadze:2003jq} that rapid classical instabilities
are present in the Fierz--Pauli theory in the {\em Minkowski
background} is unjustified.  This claim is based on the analysis of
the spatially homogeneous solutions, while to address the issue of
stability one should study dynamics of the spatially localized
excitations of finite energy. An example illustrating this point is
provided, e.g., by the massless scalar field with the negative
potential $V=-\lambda\phi^4$.  From the analysis of the spatially
homogeneous solutions one might conclude that vacuum $\phi=0$ is
perturbatively unstable in this theory, which is not the case (see,
e.g.~\cite{Son:1993ak}).}.  We will see that the instabilities of this
type are absent in our models. A physical reason is that massive
gravities with symmetry (\ref{symmetry}) can be thought of as stable
scalar theories coupled to the Einstein gravity, which is not possible
the Fierz--Pauli case.

The origin of the Boulware--Deser instability is easy to understand
within the formalism of the Goldstone fields $\phi^{\mu}$. The
Goldstone action which corresponds to Lorentz-invariant massive
gravity, including the Fierz-Pauli theory, is (cf. the second term in
Eq.~(\ref{action}))
\begin{equation}
\label{poincare}
S_G=\Lambda^4\int \sqrt{-g}d^4xF(P),
\end{equation}
where
$P=g^{\mu\nu}\eta_{\alpha\beta}\d_\mu\phi^\alpha\d_\nu\phi^\beta$ and
$\eta_{\alpha\beta}$ is a Minkowski metric. In the Minkowski
background (\ref{vacuum}), the quadratic action for the Goldstone
perturbations $\delta \phi^\mu \equiv \xi^\mu$ takes the form
\begin{equation}
L = \mu_1^2(P) (\d_\mu\xi^\alpha)^2 + \mu_2^2(P)
(\d_\mu\xi^\mu)^2, 
\label{FPlagrangian}
\end{equation}
where the coefficients $\mu_i^2(P)$ are some functions of
$P=\alpha^2-3\beta^2$ which are expressed in terms of the first and
second derivatives of the function $F(P)$. The particular expressions
are irrelevant for the argument; what is important is the fact that
the coefficients $\mu_i^2(P)$ do depend on $P$. The Lagrangian
(\ref{FPlagrangian}) describes the ghost-free theory 
only in the case
\begin{equation}
\mu_1^2(P) + \mu_2^2(P) =0\;,
\label{FPcondition}
\end{equation}
when it is proportional to $(\d_\mu\xi_\nu - \d_\nu\xi_\mu)^2$. In
general, this condition is satisfied in an isolated point
$P=P_0$. 

To see the instability, consider now the perturbations localized in
the vicinity of a given point in the background of some non-trivial
solution. In the UV limit the metric can be approximated as flat, so
the perturbations in the Goldstone sector will be described by the
Lagrangian (\ref{FPlagrangian}). However, since Goldstone fields
depend on space and time, the condition (\ref{FPcondition}) will not,
in general, be satisfied (because, for instance, the value of $P$ is
time-dependent for cosmological solutions). The Fierz-Pauli theory is
therefore UV unstable in a curved background even if this background
is locally very close to the Minkowski one.  This implies that the
cutoff scale of the low-energy effective theory in the curved
backgrounds generically is even less than in the Minkowski background,
as it should be less than a mass of the ghost mode.  A detailed
discussion of the corresponding scales in the phenomenologically
relevant backgrounds can be found in \cite{Alberto}.

Let us now repeat the same analysis for our model and show that it is
free from UV instabilities at least for backgrounds which are close to
the vacuum (\ref{vacuum}) in the UV limit.  For simplicity, consider
the model obeying the dilatation symmetry (\ref{dilatation}). The
Goldstone action has the form
\begin{equation}
\label{Zgammaaction}
S_G=\Lambda^4\int \sqrt{-g}d^4xF(Z^{ij})\;,
\end{equation}
where $Z^{ij}=X^\gamma W^{ij}$, and the quantities $X$ and $W^{ij}$
are given by Eq.~(\ref{XVYW}). Deep in the UV region where the metric
can be considered as flat, any Goldstone configuration of the form
(\ref{vacuum}) is a solution to the Goldstone field
equations\footnote{In principle, one may consider a larger class of
background, e.g. those with $\phi^i=B^i_jx^j$. For definiteness, we
restrict our discussion to rotationally invariant case.  This choice
covers, in particular, cosmological solutions obtained above.}. In
this background, the variable $Z^{ij}$ takes the values depending on
the constants $\alpha$ and $\beta$. The quadratic Lagrangian for the
Goldstone perturbations $\xi^0$ and $\xi^i$ reads
\begin{gather}
\label{piaction}
L={M_{Pl}^2}\Bigl\{ 2m_0^2(\d_0\xi_0)^2 + m_1^2(\d_i\xi_0)^2+
4m_4^2\xi_0\d_0\d_i \xi_i-m_2^2(\d_i\xi_j)^2-
(m_2^2-2m_3^2)(\d_i\xi_i)^2\Bigr\},
\end{gather}
where the kinetic coefficients $m_i$ are certain functions of $Z^{ij}$
(and therefore, of $\alpha$ and $\beta$). Their explicit expressions
are given in the Appendix B.  Using these expressions one can check
that the kinetic coefficients satisfy the constraints
\begin{eqnarray}
m_0^2 = -3\gamma m_4^2, \qquad \gamma(m_2^2-3m_3^2)=m_4^2-{1\over 2}m_1^2
\label{constraint-on-masses}
\end{eqnarray}
which follow from the symmetry (\ref{symmetry}). Note that these
constraints differ from Eqs.~(\ref{relation-for-masses}) because the
background we consider now does not correspond to the zero
energy-momentum tensor of the Goldstone fields. They reproduce
Eqs.~(\ref{relation-for-masses}) at $m_1^2=0$.

It is convenient to decompose the Goldstone perturbations into the
transverse vector $\xi^T_i$ ($\d^i\xi^T_i=0$) and two
scalars $\xi_L$ and $\xi_0$, as defined by the following relation,
\[
\xi_i = \xi^T_i + {1\over  \sqrt{-\d_i^2}}\d_i\xi_L.
\]
The Lagrangian for the vector part reads
\begin{equation}
\label{vector}
L=-M_{Pl}^2m_2^2(\d_i\xi^T_j)^2\;.
\end{equation}
Both modes in the vector sector have the dispersion relation $p_i^2=0$
and do not propagate, independently of the values of $\alpha$ and
$\beta$; this is a consequence of the symmetry (\ref{symmetry}).  There
are no instabilities in this sector.

The Lagrangian for scalar perturbations $\xi_0$ and $\xi_L$ is
\begin{gather}
L=M_{Pl}^2 \Bigl\{ 2m_0^2(\d_0\xi_0)^2+m_1^2(\d_i\xi_0)^2-
4m_4^2\xi_0\d_0\sqrt{-\d_i^2}\xi_L
-2(m_2^2-m_3^2)(\d_i\xi_L)^2\Bigr\}.
\label{largescal}
\end{gather}
In the Fourier space it can be written as 
\[
L = M_{Pl}^2 \cdot \xi^\dagger M \xi,
\]
where $\xi = (\xi_0,\xi_L)$ and the $2\times 2$ matrix $M$ has the form 
\begin{equation}
\label{M}
M= \l \begin{array}{cc}
2m_0^2\omega^2+m_1^2p^2 & -2im_4^2\omega p\\
2im_4^2\omega p& 2(m_3^2-m_2^2)p^2
\end{array} \r
\end{equation}
with $p = \sqrt{p_i^2}$. The eigenvectors of the matrix $M$ correspond
to physical excitations. The eigenvalues can be written as 
\[
M_{\pm} = {1\over 2}\Bigl\{ T \pm \sqrt{T^2-4D}\Bigr\}, 
\]
where $T={\rm Tr}(M)$ and $D={\rm det(M)}$. They determine two
dispersion relations $\omega^2_\pm(p^2)$ by the implicit equations
\[
M_{\pm} (\omega^2,p^2) =0.
\]
The system is classically stable if $\omega^2_\pm(p^2)>0$ for all
relevant $p^2$. The system has no ghosts if near the mass shell the
terms linear in $\omega^2$ are non-negative,
\begin{equation}
{\d M_{\pm}(\omega^2,p^2) \over \d \omega^2}\Bigm|_{\omega^2 =
\omega_\pm^2(p^2)} \geq 0,
\label{ghostinequality}
\end{equation}
for both modes. 

The mode which corresponds to the eigenvalue $M_-$ has the dispersion
relation $p^2=0$ and does not propagate. The inequality
(\ref{ghostinequality}) is marginally satisfied, so this mode does not
cause the UV instability. Note that the existence of the scalar mode
with the dispersion relation $p^2=0$ is guaranteed by the
reparametrization symmetry (\ref{symmetry}).  

It is worth mentioning a physical interpretation of the modes with the
dispersion relation $p^2=0$. They can be thought of as degrees of
freedom with infinite propagation velocity (unlike the ghost
condensate mode which has zero velocity at zero-derivative level and
acquires a very small velocity due to higher derivative
terms). Physically, they describe sound waves propagating through the
rigid coordinate frame selected in space by the functions
$\phi^i$. The rigidity of this frame is ensured by the symmetry
(\ref{symmetry}) and $SO(3)$ symmetry of the Goldstone action that
allow to move and rotate this frame only as a whole. Note that
infinitely fast propagating modes do not imply the violation of
causality in the absence of Lorentz invariance, but allow for
instantaneous transfer of information. A recent discussion of some of
the properties of these modes in the toy QED model can be found in
Refs.~\cite{Dvali:2005nt,Gabadadze:2004iv}.

The mode which corresponds to the eigenvalue $M_+$ has the dispersion
relation
\[
\omega^2 = v^2 p^2,
\]
where 
\begin{equation}
\label{velocity}
v^2={1\over 2}{m_1^2(m_3^2-m_2^2)\over m_4^4-m_0^2(m_3^2-m^2_2)}.
\end{equation}
The absence of classical instabilities thus requires 
\begin{equation}
{m_1^2(m_3^2-m_2^2)\over m_4^4-m_0^2(m_3^2-m^2_2)}>0.
\label{v2>0}
\end{equation}
When this condition is satisfied, Eq.~(\ref{ghostinequality}) which
ensures the absence of ghosts translates into the following inequality
(see Appendix B for details),
\begin{equation}
m_0^2-{m_4^4\over m_3^2-m_2^2}>0,
\label{noghostcondition}
\end{equation}
in agreement with the result of Ref.~\cite{Dubovsky:2004sg}. Thus,
there are neither classical instabilities nor ghosts in our model
provided that both conditions (\ref{v2>0}) and
(\ref{noghostcondition}) are satisfied. These conditions are
compatible with the constraints (\ref{constraint-on-masses}).

For a flat background $m_1^2=0$, so one may worry about UV stability
of an arbitrarily close background with the positive value of $m_1^2$
and, consequently, negative velocity $v^2<0$. In the vicinity of the
point where $v^2 =0$ the higher-derivative terms in the dispersion
relation become important, so that it takes the form $\omega^2 = v^2
p^2 + \alpha p^4/\Lambda^2$, where $\alpha$ is a coefficient of order
one which we assume to be positive. It is clear now that close to the
point $v^2=0$ the instability occurs only at very low momenta, i.e.,
in the IR region.  The situation here is the same as in the ghost
condensate model. By analogy we expect that accounting for mixing with
gravity for higher derivative terms leads to the IR instability of
this type already in the flat background with $m_1^2=0$.

The case $\gamma=0$ when the Goldstone action depends only on $W^{ij}$
requires a separate consideration.  Using the expressions for the
graviton masses given in the Appendix A, Eqs.~(\ref{m0})--(\ref{m4}),
one finds that in this case $m_0^2=m_1^2=m_4^2=0$ in the Minkowski
background. Therefore, this is a theory with $\Delta=0$ discussed in
the end of Sect.~\ref{sect:linear}.  At the one-derivative level this
theory possesses a symmetry $\phi^0\to\phi^0+ \xi^0(\phi^0)$ apart
from the symmetry $\phi^i\to\phi^i+ \xi^i(\phi^0)$. (This symmetry
should be imposed at the higher-derivative level in order to avoid
ghosts.)  Naively, one may expect that the above symmetries imply that
all modes should always have the dispersion relation $p^2=0$.
However, the situation is more subtle. From Eq.~(\ref{coeffs}) of
Appendix B one finds that in the curved background only the mass
$m_0^2$ is equal to zero, while $m_1^2$ and $m_4^2$ may be non-zero.
As a result, in curved backgrounds, in addition to two solutions with
$p^2=0$, $\xi_L(t)$ and $\xi_0(t)$, there is also a mode with the
velocity
\begin{equation}
\label{velocity1}
v^2={1\over 2}{m_1^2(m_3^2-m_2^2)\over m_4^4}\;,
\end{equation}
which is very large for backgrounds close to the Minkowski one because
$m_1^2\sim m_4^2$ are both very small.  Of these three modes, only one
($\xi_L(t)$) is seen in the quadratic action about flat background at
the one-derivative level.  We believe that further analysis is needed
to understand whether the two new modes lead to the problems like low
strong coupling scale.  Note, however, that unlike in the Fierz--Pauli
case, the new modes are not ghosts, provided the condition
(\ref{noghostcondition}) holds.

We see that the situation in our models is quite different from that
in the Fierz-Pauli theory of massive gravity. Unlike the latter, our
models are free of ghosts in a finite region of coefficients $m_a^2$
in Eq.~(\ref{piaction}) (and, therefore, of constants $\alpha$ and
$\beta$) which includes the point corresponding to the flat background
with $m_1^2=0$. 
Thus, with a proper choice of the function $F$ and
higher derivative terms, our models are UV stable at least for
backgrounds close to the flat one. The Boulware-Deser instability is
absent.

\section{Discussion}
\label{discussion}
In order to be more than a theoretical exercise, the theory of
Lorentz-violating massive gravity must eventually address the
fundamental puzzles of modern cosmology such as the origin of dark
matter and dark energy. The class of models discussed in this paper provides
a number of possibilities in this direction.

\begin{figure}
\epsfig{file=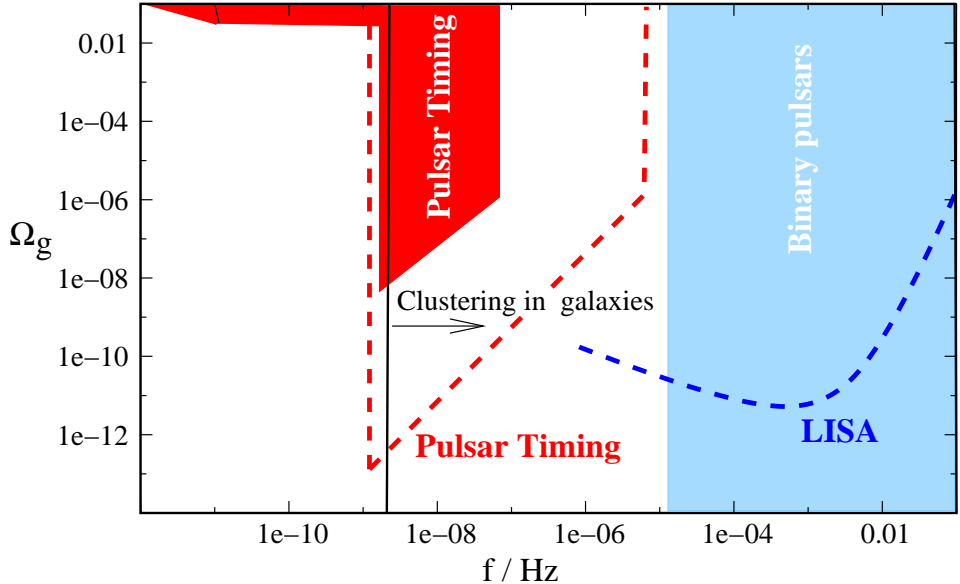}
\caption{Limits on the gravitational wave signal in the
frequency/relative graviton abundance plane. Light shaded region is
excluded by the observations of binary pulsars~\cite{pulsars}). Dark
shaded region is excluded by the timing of the millisecond
pulsars~\cite{timing}.  Dashed lines show the expected sensitivity of
the Australian pulsar timing array and LISA. Frequencies higher than
that marked by the solid line correspond to graviton masses large
enough to allow for gravitons to cluster in galaxies. Note, that if
all of the galactic dark matter is comprised of massive gravitons then
the gravitational wave signal corresponds to graviton abundance
$\Omega_g\sim 10^5$.  }
\label{fig1}
\end{figure}

As follows from Sect.~\ref{cosmology}, the evolution of the Universe
may naturally lead to the attractor which corresponds to the theory
possessing the dilatation symmetry (\ref{dilatation}). In this case,
the relations (\ref{relation-for-masses}) among masses imply that the
growing term in the Newtonian potential vanishes. Even in this simplest
version, the model has a number of features interesting from
the cosmological and observational  points of view. 
First, the massive graviton itself is a
candidate for the dark matter particle \cite{Dubovsky:2004ud}. This
possibility is observationally testable, the current limits being
plotted in Fig.~\ref{fig1}.
The constraints will be improved in the near future by the data from
the Australian pulsar timing array~\cite{Hobbs:2004uc}. If massive
gravitons do not constitute all of the dark matter, they are still
detectable in a certain range of masses because they would produce a
unique monochromatic signal in a gravitational wave detectors such as
LISA~\cite{Bender:2003uv}.

Second, the Goldstone fields give two extra contributions to the
Friedmann equation (\ref{00eq}) which we denoted $\rho_1$ and $\rho_2$
in Sect.~\ref{cosmology}. For $1/3<\gamma<1$ 
the first of these contributions $\rho_1$
behaves like a ``quintessence'' with the equation of state varying
from $w=-1$ to $w=-1/3$ for different values of the parameter $\gamma$, which
characterizes the dilatation symmetry emerging in the cosmological 
attractor (see Eq.~(\ref{dilatation})). The second
contribution $\rho_2$ behaves as a cosmological constant. 

An interesting special situation takes place for $\gamma=1/3$.  In this case
both contributions $\rho_1$ and $\rho_2$ have vacuum equation of state
$\omega=-1$. As a result the acceleration rate of the late de Sitter phase is
a dynamical quantity, determined by the initial conditions in the Goldstone
sector rather than by parameters of the action.  This is similar
to situation in the unimodular gravity\footnote{We thank John March-Russel
for pointing out this similarity.} \cite{Weinberg:1988cp}, where cosmological
constant is also a constant of integration. This similarity appears to be not
just a coincidence. The metric determinant $g$ is invariant under the symmetry
(\ref{symmetry}) and under the dilatation symmetry (\ref{dilatation}) with
$\gamma=1/3$, so the symmetry group of massive gravity is a subgroup of the
unimodular gravity in this case.

On the other hand there is an important difference between massive
gravity with $\gamma=1/3$ and unimodular gravity. Namely, the only
difference between unimodular gravity and the Einstein theory (at
least at the classical level) is that in the former case solutions
with arbitrary values of cosmological constant are present
independently of the value of the vacuum energy. On the contrary, in
massive gravity the contribution of the Goldstone sector has the form
of a cosmological constant only for flat homogeneous cosmological
solutions. In particular, the initial conditions in the Goldstone
sector may vary in space resulting in solutions with different values
of the acceleration rate in different parts of the Universe. Such
solutions are absent in both general relativity and unimodular
gravity.  If nothing else, this allows the application of the
anthropic arguments~\cite{Weinberg:1988cp}.  There is a caveat,
however. In order to cancel a bare cosmological constant that is much
larger than $\Lambda^4$ (where $\Lambda$ is the cutoff scale of our
model) one needs a fine-tuning to keep the mass of the graviton from
being too large. It is not impossible to imagine an anthropic
explanation for this fine-tuning as well --- according to the
estimates of Ref.~\cite{Dubovsky:2004ud} relic massive gravitons with
masses higher than $(10^{15}{\mbox{cm}})^{-1}$ (upper bound from the
timing of the binary pulsars~\cite{pulsars}) are likely to overclose
Universe.  Alternatively, one may hope that the unusual properties of
cosmological solutions in massive gravity may be a first step towards
a dynamical solution of the dark energy problem.

The situation may become more complicated if the model is kept away
from the dilatation-symmetric attractor by a fine-tuning of the
cosmological evolution or some other mechanism (e.g., if
$X^\gamma/a^2$ is proportional to some power of $\log a$ at late
times, instead of being a constant). Then the potential of a
point-like source acquires --- formally --- the linearly growing
contribution, Eq.~(\ref{confpotential}). There are two distance scales
associated with this contribution. The first one, $l_1 \sim 1/m$,
determines the distances where the growing term starts to dominate
over the conventional one. The second scale is $l_2 =(M_{Pl}^2/M
m^2)$; it depends on the mass $M$ of the source. At distance $l_2$ the
potential $\Phi$ becomes of order unity indicating the breakdown of
perturbation theory and possible onset of a non-linear regime.  Note
that in the gauge we are using, the metric components that become
large are
\[
h_{0i}\sim n_i{t\over l_2}\;,
\]
where $n_i$ is a unit vector in the direction of the source.
Consequently, non-linear regime starts at the moment of time $t_2\sim l_2$
rather than at a certain distance from the source.

To understand this qualitatively, recall that our choice of the
integration constant $\psi_0(x^i)=0$ corresponds to the initially
homogeneous Universe. One can view the Goldstone sector as a
(multi-component) fluid which is accreted by sources after they are
formed. Eventually, this accretion results in the onset of the
non-linear regime; qualitatively, this happens at the time of order
$l_2$ after the formation of sources.  It is tempting to speculate
that this non-linear phase may result in the non-zero profile of the
ghost condensate mode $\psi_0$ such that the linearly growing term in
the potential (\ref{Phi}) is canceled. The corresponding non-linear
dynamics is presumably similar to that of the ghost condensate model
and is not sufficiently understood at the moment (see
\cite{Frolov:2004vm,Krotov:2004if,Mukohyama:2005rw} for some proposals
in this direction).  One characteristic feature of the ghost
condensate dynamics is the presence of strong retardation
effects~\cite{Arkani-Hamed:2003uy,Dubovsky:2004qe,Peloso:2004ut}, so
one may think that the cancellation is incomplete, leading to the
logarithmically growing potential needed to explain flat rotation
curves.  Note that non-linear effects related to the ghost condensate
mode are present even when linearly growing terms in the potential are
forbidden by dilatation symmetries, so understanding of these effects
is one of the most pressing questions for this kind of models.
\section*{Acknowledgments} 
We thank Nima Arkani-Hamed, Thomas Gregoire,
Markus Luty, John March-Russel, Alberto Nicolis, Riccardo Rattazzi,
Valery Rubakov, 
Mikhail Sazhin, Matt Schwartz and Sergei Sibiryakov
for useful discussions and correspondence. 
SD thanks Harvard Theory Group
where part of this work was done for a warm
hospitality. The work of P.T. is supported by IISN, Belgian Science
Policy (under contract IAP V/27).
\section*{Appendix A}

In this Appendix we calculate mass terms of the gravitational field in
the Friedmann background.  The mass terms come from the expansion of
the second term in the r.h.s. of Eq.~(\ref{action}) in powers of the
perturbation $\delta g_{\mu\nu}$ to the quadratic order about
background metric
\begin{equation}
ds^2 = g_{\mu\nu}\, dx^\mu dx^\nu = a^2(\eta)(d\eta^2 - d{\bf x}^2 ) \; .
\label{FRW}
\end{equation} 
With the definitions (\ref{XVYW}) one has
\begin{eqnarray}\nonumber
\sqrt{-(g+\delta g)} &=& a^4 + 
\frac{a^2}{2}\left( \delta g_{00} - \delta g_{ii}\right) 
- {1\over 8} \delta g_{00}^2 
- {1\over 4} \delta g_{00} \delta g_{ii} + {1\over 8} \delta g^2_{ii}
+ \half \delta g_{0i}^{2} - {1\over 4} \delta g_{ij}^2+ \ldots,\\ \nonumber
X(g+\delta g) &=& X(g)\, \left[1 - \frac{1}{a^2} \delta g_{00} + \frac{1}{a^4}
\left( \delta g_{00}^2 - \delta  g_{0i}^2\right) +\ldots \right],
\\\nonumber
W^{ij}(g+\delta g) &=& W(g) \left[-\delta_{ij} - \frac{1}{a^2}\delta g_{ij} - 
\frac{1}{a^4} \delta g_{ik} \delta g_{kj} +\ldots \right] \; ,
\end{eqnarray}
where
\begin{equation}
W \equiv -\, \frac{1}{3} \delta_{ij} W^{ij} \; . 
\label{Wdef}
\end{equation} 

Due to the rotational symmetry, the derivatives of $F$ up to the
second order are expressed in terms of the 6 scalar quantities $F_X$,
$F_W$, $F_{XX}$, $F_{XW}$, $F_{WW1}$ and $F_{WW2}$ which are defined
as follows,
\begin{eqnarray}
{\d \over \d X} F(X,W^{ij})&=& F_X ,\\
{\d \over \d W^{ij}} F(X,W^{ij}) &=& F_W \delta_{ij},\\
{\d^2 \over \d X^2} F(X,W^{ij})&=& F_{XX},\\
{\d^2 \over \d X \d W^{ij}} F(X,W^{ij}) &=& F_{XW} \delta_{ij},\\
{\d^2 \over \d W^{ij} \d W^{mn}} F(X,W^{ij}) &=& F_{WW1}
\delta_{ij}\delta_{mn} + F_{WW2} (\delta_{im}\delta_{jn} 
+ \delta_{in}\delta_{jm}).
\end{eqnarray}
The derivatives on the l.h.s. of these equations are all evaluated at
the point $X(g)$, $W^{ij}(g)$. 

With these definitions, the linear contribution to the expansion of
the second term in Eq.~(\ref{action}) is
\begin{eqnarray}
\nonumber
a^2 \left ({1\over 2} F - X F_X \right) \delta g_{00} -
a^2 \left ({1\over 2} F + W F_W \right) \delta g_{ii} 
\end{eqnarray}
The corresponding Friedmann equations are
\begin{eqnarray}
\frac{3a'^{\, 2}}{a^4} &=& 8\pi G \left[\Lambda^4 \left(2X F_X - F\right) +
\rho_m \right] \; , \nonumber \\
\frac{2a''}{a^3}-\frac{a'^{\,2}}{a^4} &=& -\; 8\pi G \left[\Lambda^4 
\left(2W F_W +F\right) + p_m \right] \; ,
\end{eqnarray}
where $\rho_m$, $p_m$ are the energy density and pressure of
matter. Combination of these two equations gives the equation of motion
for the field $\phi^0$,
\begin{equation}
\label{phi0_A}
a^3\, \sqrt{X}\, F_X = {\rm const}\;.
\end{equation}

In the Minkowski background one has
\begin{eqnarray}
\label{constraint1}
&&F - 2 X F_X =0,\\
\label{constraint2}
&&F + 2 W F_W =0.
\end{eqnarray}
For a generic function $F$ these equations are satisfied for some $X$, $W$.

The quadratic part of the Lagrangian Eq.~(\ref{action}) with respect to metric
perturbations about Friedmann background is
\begin{gather}
\nonumber
\Lambda^4 \Bigl\{{1\over 2} X^2 F_{XX}+ {1\over 2} X F_{X} - {1\over 8} F  
\Bigr\} \delta g_{00}^2 + 
\Lambda^4 \Bigl\{{1\over 2} F -  X F_{X}  
\Bigr\} \delta g_{0i}^2 \\
+\Lambda^4 \Bigl\{ XW F_{XW} - {1\over 2}W F_{W} + {1\over 2}X F_{X}  
- {1\over 4} F \Bigr\} \delta
g_{00} \delta g_{ii} \nonumber \\
+ \Lambda^4 \Bigl\{{1\over 2} W^2 F_{WW1} + {1\over 2} W F_{W} + {1\over 8} F 
\Bigr\} \delta g_{ii}^2
+ \Lambda^4 \Bigl\{ W^2 F_{WW2} - W F_{W} - {1\over 4} F 
\Bigr\} \delta g_{ij}^2,
\end{gather}

Comparing this expression to Eq.~(\ref{masterm1}) 
one finds for the masses of the gravitational field in the Minkowski vacuum,
\begin{eqnarray}
\label{m0}
m_0^2 &=& {\Lambda^4\over M_{Pl}^2} 
\Bigl\{ X F_X + 2X^2 F_{XX}\Bigr\},\\
\label{m1}
m_1^2 &=& 0,\\
\label{m2}
m_2^2 &=& - {\Lambda^4\over M_{Pl}^2} 
\Bigl\{ 2XF_X + 4 W^2 F_{WW2} \Bigr\},\\
\label{m3}
m_3^2 &=&{\Lambda^4\over M_{Pl}^2} 
\Bigl\{ - X F_X + 2 W^2 F_{WW1} \Bigr\} ,\\
\label{m4}
m_4^2 &=&-{\Lambda^4\over M_{Pl}^2}
\Bigl\{ X F_X + 2 X W F_{XW} \Bigr\}.
\end{eqnarray}

\section*{Appendix B}
In this Appendix we provide some intermediate formulas skipped
in the section~\ref{stability}.

The explicit expressions for the kinetic coefficients in the
action (\ref{piaction}) are
\begin{eqnarray}
\nonumber
m_0^2&=&-{6\Lambda^4\over M_{Pl}^2}\Bigl\{\gamma(\gamma-{1\over 2})F_ZZ
-3\gamma^2 F_{ZZ1}Z^2-2\gamma^2 F_{ZZ2} Z^2\Bigr\},\\
\nonumber
m_1^2&=&{2\Lambda^4\over M_{Pl}^2}(3\gamma-1)F_ZZ,\\
\nonumber
m_2^2&=&{2\Lambda^4\over M_{Pl}^2}\Bigl\{F_ZZ-2F_{ZZ2}Z^2\Bigr\},\\
\nonumber
m_3^2&=&{2\Lambda^4\over M_{Pl}^2}\Bigl\{{1\over 2}F_ZZ+F_{ZZ1}Z^2\Bigr\},\\
m_4^2&=&{2\Lambda^4\over M_{Pl}^2}\Bigl\{(\gamma-{1\over 2})F_ZZ-3\gamma F_{ZZ1}Z^2-2\gamma F_{ZZ2}Z^2\Bigr\}.
\label{coeffs}
\end{eqnarray}
Here $Z\equiv-Z^{ij}\delta_{ij}/3$, while the scalar functions
$F_Z$, $F_{ZZ1}$ and $F_{ZZ2}$ are defined by the following relations,
\begin{eqnarray}
\nonumber
{\d \over \d Z^{ij}} F(Z^{ij}) &=& F_Z \delta_{ij},\\
\nonumber
{\d^2 \over \d Z^{ij} \d Z^{mn}} F(Z^{ij}) &=& F_{ZZ1}
\delta_{ij}\delta_{mn} + F_{ZZ2} (\delta_{im}\delta_{jn} 
+ \delta_{in}\delta_{jm}).
\end{eqnarray}
It is straightforward to check that coefficients (\ref{coeffs}) satisfy 
relations (\ref{constraint-on-masses}).

Calculation of the no-ghost condition (\ref{noghostcondition}) proceeds as
follows
\begin{equation}
\label{derivative}
{\d M_{\pm}(\omega^2,p^2) \over \d \omega^2}\Bigm|_{\omega^2 =
\omega_\pm^2(p^2)}={1\over 2}\l T'\pm{TT'-2D'\over |T|}\r={D'\over T}\;,
\end{equation}
where prime denotes differentiation with respect to $\omega^2$, and
in algebraic transformation we were taking into account, 
that we are taking the derivative of the eigenvalue which is zero on-shell.
Plugging explicit expressions for $D$ and $T$, following from Eq.~(\ref{M}),
we obtain the following condition for the propagating
mode to be not a ghost
\begin{equation}
\label{notaghost}
{m_0^2(m_3^2-m_2^2)-m_4^4\over 2m_0^2v^2+m_1^2+2(m_3^2-m_2^2)}>0\;,
\end{equation}
where $v^2$ is given by Eq.~(\ref{velocity}). Using explicit expression 
(\ref{velocity}) one can check that at $v^2>0$ this condition is equivalent
to (\ref{noghostcondition}).

\end{document}